\documentclass[12pt]{article}
\usepackage{epsfig}
\usepackage{graphicx}
\usepackage{./pazha}
\tightenlines
\def\a{$^{\mbox{\small a}}$}
\def\b{$^{\mbox{\small b}}$}
\def\c{$^{\mbox{\small c}}$}
\def\d{$^{\mbox{\small d}}$}
\def\e{$^{\mbox{\small e}}$}
\def\f{$^{\mbox{\small f}}$}
\def\deg{^\circ}
\def\*{$^{*}$}
\def\INTEGRAL{\hbox{INTEGRAL}}
\sloppypar

\begin{document}
\baselineskip 21pt
{\sl to be published in Astronomy Letters, v. 33, n. 12, in press (2007)\\
translated from Pis'ma v Astron. Zhurn., v. 33, n. 12, p. 883 (2007)}
\vspace{2cm}

\title {\bf COSMIC GAMMA-RAY BURST 060428C DETECTED IN THE
FIELDS OF VIEW OF THE {\it {IBIS}\/} AND {\it {SPI}\/} TELESCOPES OF
THE {\it {INTEGRAL}\/} OBSERVATORY AND ITS EARLY AFTERGLOW}

\author
{\bf \hspace{-1.3cm}\copyright\, 2007\ \ 
S.A.\,Grebenev\affilmark{*}, I.V.\,Chelovekov}

\affil{{\sl Space Research Institute, Russian Academy of
    Sciences}\\ [-1mm]
Profsoyuznaya 84/32, Moscow 117997, Russia}
\vspace{2mm}
\centerline{\rm Submitted on July 10, 2007}

\vspace{2mm}
\baselineskip 15.2pt
\noindent
Analyzing archival data of the \INTEGRAL\ observatory we
discovered and localized a cosmic $\gamma$-ray burst detected
within fields of view of the IBIS/ISGRI and SPI telescopes on
April 28, 2006. The burst has not been revealed by the INTEGRAL
Burst Alert System (IBAS), so the information on its coordinates
has not been distributed in time and search for the afterglow
has not been carried out. The burst has been also detected by
the KONUS/WIND and RHESSI spacecrafts. Its fluence was
$2.3\times 10^{-6}$ erg cm$^{-2}$ in the 20--200 keV energy
band, the maximum flux was $3.6\times10^{-7}$ erg cm$^{-2}$
s$^{-1}$ ($3.9$ phot cm$^{-2}$ s$^{-1}$). The burst had a
complicated multi-peak profile and was outstanding of the
typical bursts by increasing its spectral hardness with
time. The emission spectrum near the flux maximum was
characterized by the photon index $\alpha\simeq-1.5$ and the
peak energy $E_p\simeq95$ keV. The burst lasted $\sim12$ s, then
we observed its afterglow at energies 15--45 keV decaying
according to a power law with the index $\gamma\sim-4.5$. The
spectral hardness decreased substantially during the afterglow.



\vfill

\noindent\rule{8cm}{1pt}\\
{$^*$ e-mail address $<${\sl sergei@hea.iki.rssi.ru}$>$}

\clearpage

\section*{INTRODUCTION}

There were about 50 cosmic $\gamma$-ray bursts detected within
fields of view of main telescopes of the INTEGRAL observatory
during its 5 years of in-orbit operation. Most of them were
discovered and then (within 30--200 s) localized with the IBAS
automatic system (Mereghetti et al. 2003) enabling quick
reorientation of X-ray, optical and radio telescopes around the
world towards sources of the bursts and, in many cases, detection
of their afterglow (see table of the detected bursts at
http://ibas.iasf-milano.inaf.it).

In this paper we report the discovery in the INTEGRAL archival
data of a $\gamma$-ray burst which was not revealed by the IBAS
system. The burst was detected on April 28, 2006 at
02\uh30\um35\us\ UT with the IBIS gamma-ray telescope (the
ISGRI detector) and SPI gamma-ray spectrometer. We assigned
the name GRB\,060428C for this burst taking into account that
two more bursts were detected later during the same day by the SWIFT
orbital observatory (Mangano et al. 2006; Campana et al. 2006),
referred in GCN telegrams as GRB\,060428A and GRB\,060428B.

Apart the INTEGRAL observatory, GRB\,060428C was also detected by
the RHESSI
(http://grb.web.psi.ch/grb\underline{~}list\underline{~}2006.html)
and WIND (its KONUS detector, see event 2730 at
http://gcn.gsfc.nasa.gov/konus\underline{~}grbs.html)
spacecrafts, with the trigger occurred only at the final stage of
the burst in the latter case. These spacecrafts were unable to
localize a source of the burst, therefore information on these
detections has not been distributed and search for the afterglow has
not been carried out.

The results of temporal and spectral analyses of the INTEGRAL
data obtained during the observations of GRB\,060428C are
presented below. The burst was modestly long (lasted less than
$15$ s) and intense (the fluence has reached $2.3\times10^{-6}\
\mbox{erg cm}^{-2}$ in the 20--200 keV band) that allowed us to
investigate its early X-ray afterglow in addition to the main
event.

\section*{INSTRUMENTS AND DATA ANALYSIS}

{\sl The International Gamma-Ray Astrophysics Laboratory INTEGRAL\/}
(Winkler et al. 2003) was placed into a high-apogee orbit by the
PROTON launcher on October 17, 2002 (Eismont et al. 2003). There
are four instruments on its board designed for simultaneous
observations in $\gamma$-ray, X-ray and optical bands. In this
paper we use data from the ISGRI detector (Lebrun et al. 2003)
of the IBIS gamma-ray telescope (Ubertini et al. 2003) and
from the SPI gamma-ray spectrometer (Vedrenne et
al. 2003). GRB\,060428C occurred $\sim9\deg$ away from the
target INTEGRAL was pointed to, so it was outside the fields of
view of the JEM-X X-ray monitor and the OMC optical monitor
which are
narrower than the fields of view of the main telescopes.

The IBIS telescope uses the principle of coded aperture to
reconstruct images of the sky and study properties of individual
sources. The telescope has a field of view of $30\deg \times 30
\deg$ ($9\deg\times 9\deg$ of it is fully coded) and an angular
resolution of 12\arcmin\ (FWHM). Such a resolution allows one to
determine the position of bright sources with an accuracy better
than 2\arcmin. The ISGRI detector of this telescope consists of
128$\times$128 CdTe semiconductor elements, with maximum
sensitivity in the 18--200 keV energy band. The detector's total
area is equal to 2620 cm$^2$, the effective area for sources
near the center of the field of view is $\sim 1100$ cm$^2$ (half
of the detector is shaded by opaque mask elements).

SPI spectrometer works also using the principle of coded
aperture, but has the lower angular resolution $2\fdg8$
(FWHM). The external diameter of the hexagonal field of view is
equal to $35\deg$\ (of the fully coded region --- $16\deg$). The
positional sensitivity is provided by 17 cryogenic Germanium
detectors with a total area of $\sim450$ cm$^2$, detecting
photons in the 20 keV -- 8 MeV energy band. The instrument's
effective area is $\sim2$ times smaller than the geometrical one
due to shading by the mask.

We found GRB\,060428C in the course of our studies under the
project dedicated to searching for X-ray bursts in open (public)
IBIS/ISGRI data. Methods of data reduction and analysis used in
this project are described in detail by Chelovekov et
al. (2006, 2007). They are based on the standard package (OSA)
of procedures for scientific analysis of the INTEGRAL data. The
comprehensive analysis of burst's properties presented in this
paper is carried out using data reduction procedures developed
for IBIS/ISGRI at the Space Research Institute (Moscow). The
general description of these procedures can be found in the
paper by Revnivtsev et al. (2004).  For the spectral analysis we
used the standard response matrix of the OSA package (rmf-file
of version 12 and arf-file of version 6), which has made good
showing being applied to the spectral approximation of the Crab
nebula. The spectrum of the nebula was adopted to be
$dN(E)/dE=10\,E^{-2.1}$ phot cm$^{-2}$ s$^{-1}$ keV$^{-1}$,
where $E$ is energy in keV. To perform spectral analysis we used
the XSPEC software package.

The data analysis for the SPI gamma-ray spectrometer has been
completely based on the OSA package, but as long as we studied a 
$\gamma$-ray burst, the background information has been
determined from two time intervals immediately before and just
after the burst. Reconstructing the broad-band spectrum we
first fitted a normalization factor of the SPI data relatively
the IBIS/ISGRI data together with parameters of the spectral
model, and then frozen it at the obtained value (0.8).

\section*{TIME PROFILE OF THE BURST}

\begin{figure}[ht]
\centering
\epsfig{file=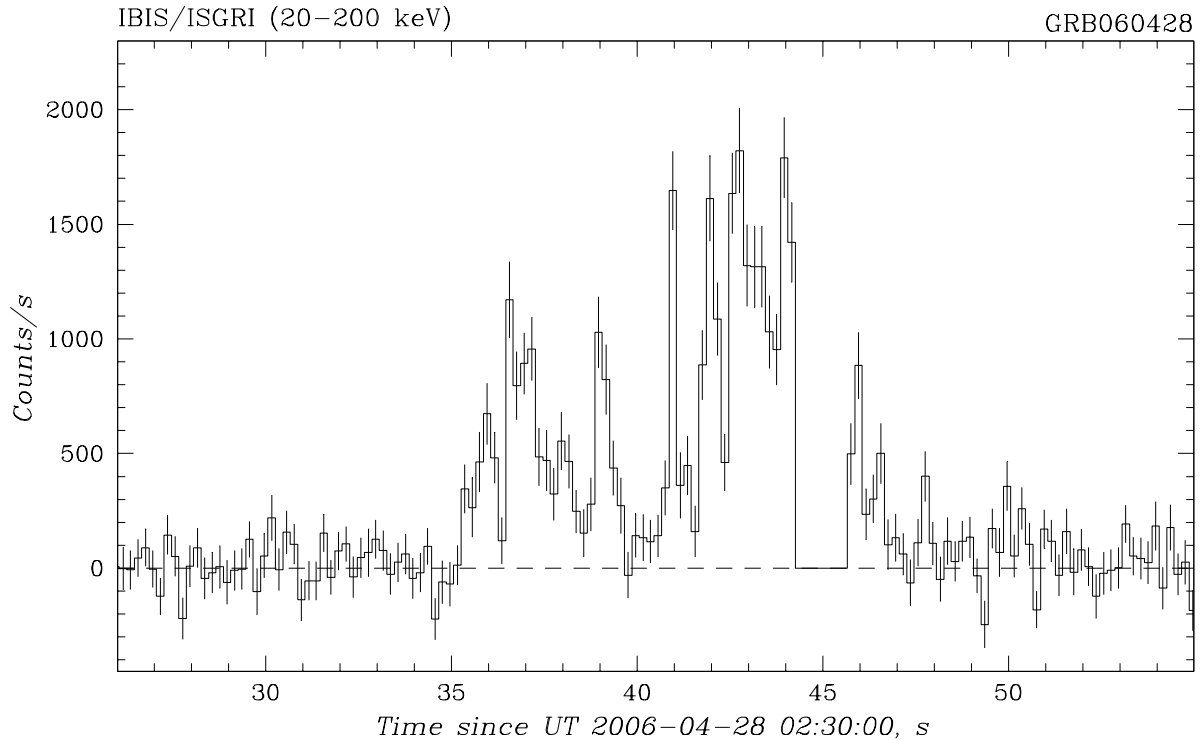, width=0.98\linewidth}

\caption{\rm Time profile of GRB\,060428C obtained by the
IBIS/ISGRI telescope in the 20--200 keV energy band. Time on the
X-axis is given in seconds after 2\uh30\um00\us\ UT, the time
bin is 0.2 s.
\label{fig:lc_isgri_0.2s}}
\end{figure}

Figure \,{\ref{fig:lc_isgri_0.2s}} shows a profile of the burst,
measured with the IBIS/ISGRI telescope in the broad 20--200 keV
energy band with a time resolution of 0.2 s. Time on the X-axis
starts since 2\uh30\um00\us\ UT April 28, 2006. Each point in
this figure is obtained by direct reconstruction of the image of
the sky in the field of view of the telescope with subtraction
of the background and subsequent measurement of the flux from
the burst's source. The count rate was corrected for dead-time
effects and efficiency of observations (the effective area
corresponding to a source position within the field of view of
the telescope). A drop in the light curve at 44--46 s arose due
to the overflow of a telemetry buffer (which is downloaded every 8
s).

The burst started at $t_0=2$\uh30\um35\us\ UT and lasted 
$\sim12$~s. The figure shows that after the end of this (main)
event, the count rate did not returned immediately to the zero
level but exceeded it during some period of time. It means that
the telescope has detected a weak remaining emission
(afterglow). The profile of the main event had a complicated
time structure, consisting of many peaks of different amplitude
and duration (down to 0.1 s). There are two comparatively long
flares, separated by a drop of 1.5 s, clearly distinguished in
the profile. The maximum flux, $(3.6\pm0.3)\times10^{-7}$ erg
cm$^{-2}$ s$^{-1}$ (or $\sim 3.9$ phot cm$^{-2}$ s$^{-1}$) in
the 20--200 keV band, was reached at 2\uh30\um42\fs6\ UT. This
value was determined using a 1-s time interval near the maximum.

\begin{figure}[hp]
\centering
\epsfig{file=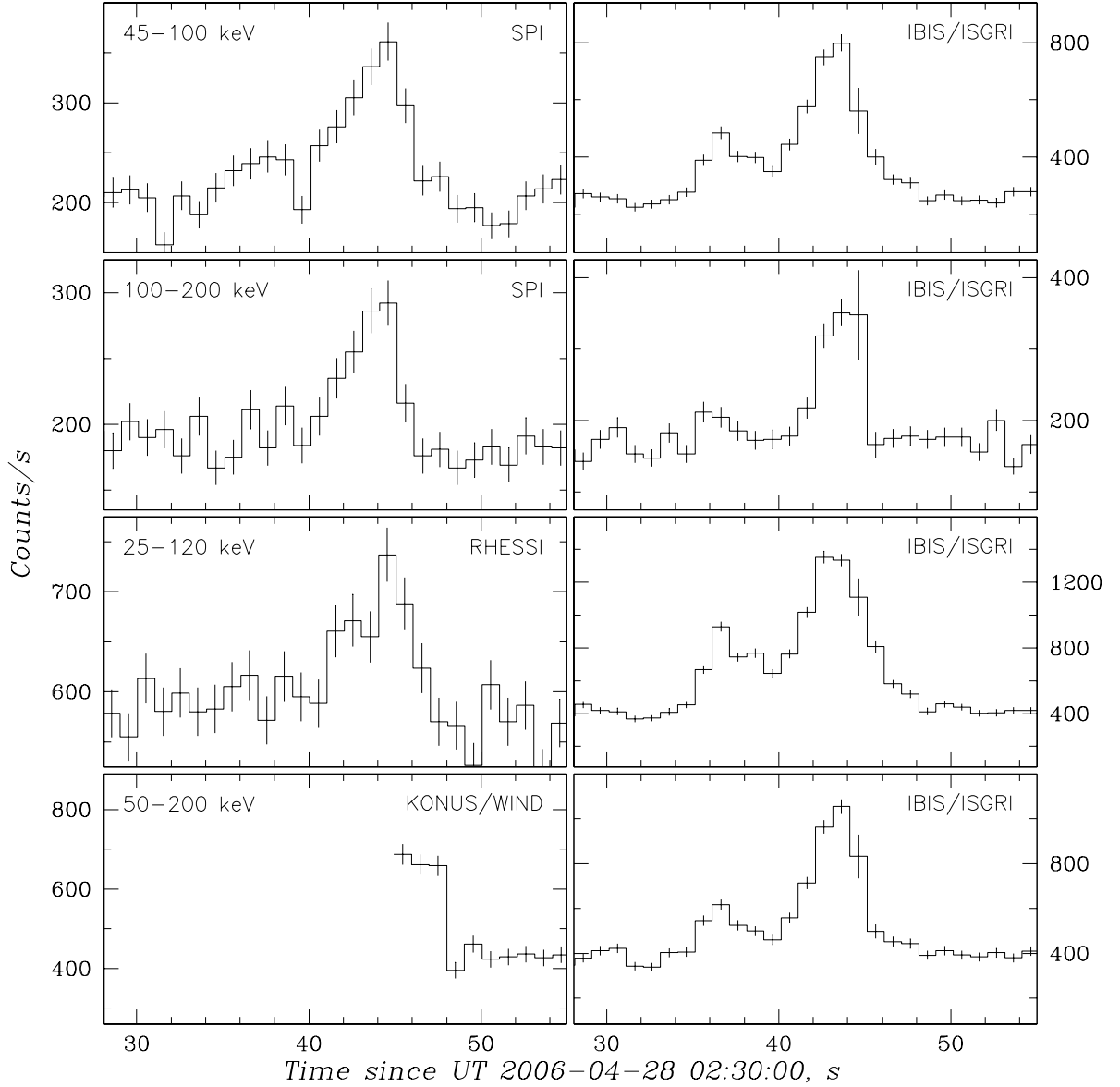,width=0.98\linewidth}

\caption{\rm Comparison of time records of the count rate during
GRB\,060428C by various instruments observed the burst. Time on
the X-axis is given in seconds after 2\uh30\um00\us\ UT, the
time bin is 1 s. The background count rate was not subtracted.}
\label{fig:lc_others}
\end{figure}

Time profiles of the burst (time records of the count rate with
a resolution of 1 s) obtained by the SPI spectrometer of the
INTEGRAL observatory in the 45--100 and 100--200 keV bands, and
by the spacecrafts RHESSI and KONUS/WIND in the 25--120 keV and
50--200 keV bands respectively are shown in the left column of
Fig.\,\ref{fig:lc_others}. The background count rate was not
subtracted here. The similar time records of the IBIS/ISGRI count
rate taken in the same energy bands are shown in the right
column of the figure, for comparison. Due to different designs
of the detectors in these experiments (having different
dependences of sensitivity on energy) the usage of the same
energy bands does not ensure the complete identity of the
measured light curves. Nevertheless the figure shows a good
enough qualitative coincidence of the curves obtained by
different instruments.  In the cases of RHESSI and, especially,
KONUS/WIND the figure indicates a notable delay in detection
of the burst relatively its detection by IBIS/ISGRI. The delay
is connected with large distances between these spacecrafts
requiring some extra time for a signal to travel from one
spacecraft to another.

Producing the IBIS/ISGRI light curve we reconstructed the drop
in the count rate during the interval of 44--46 s connected with
the overflow of the telemetry buffer. We used the fact that
information from some small parts of the detector has continued to come
during the drop. 
The accuracy of such a procedure is confirming by a comparison
of the resulted burst profiles with the data of other instruments. In
general, the figure shows that statistical significance of
observations of this burst by the IBIS/ISGRI telescope
substantially exceeds the abilities of the other instruments. In
particular, no one of the instruments has been able to confidently
resolve the first long flare in the time profile of the burst
relatively the more intense second flare.

\begin{figure}[th]
\centering
\epsfig{file=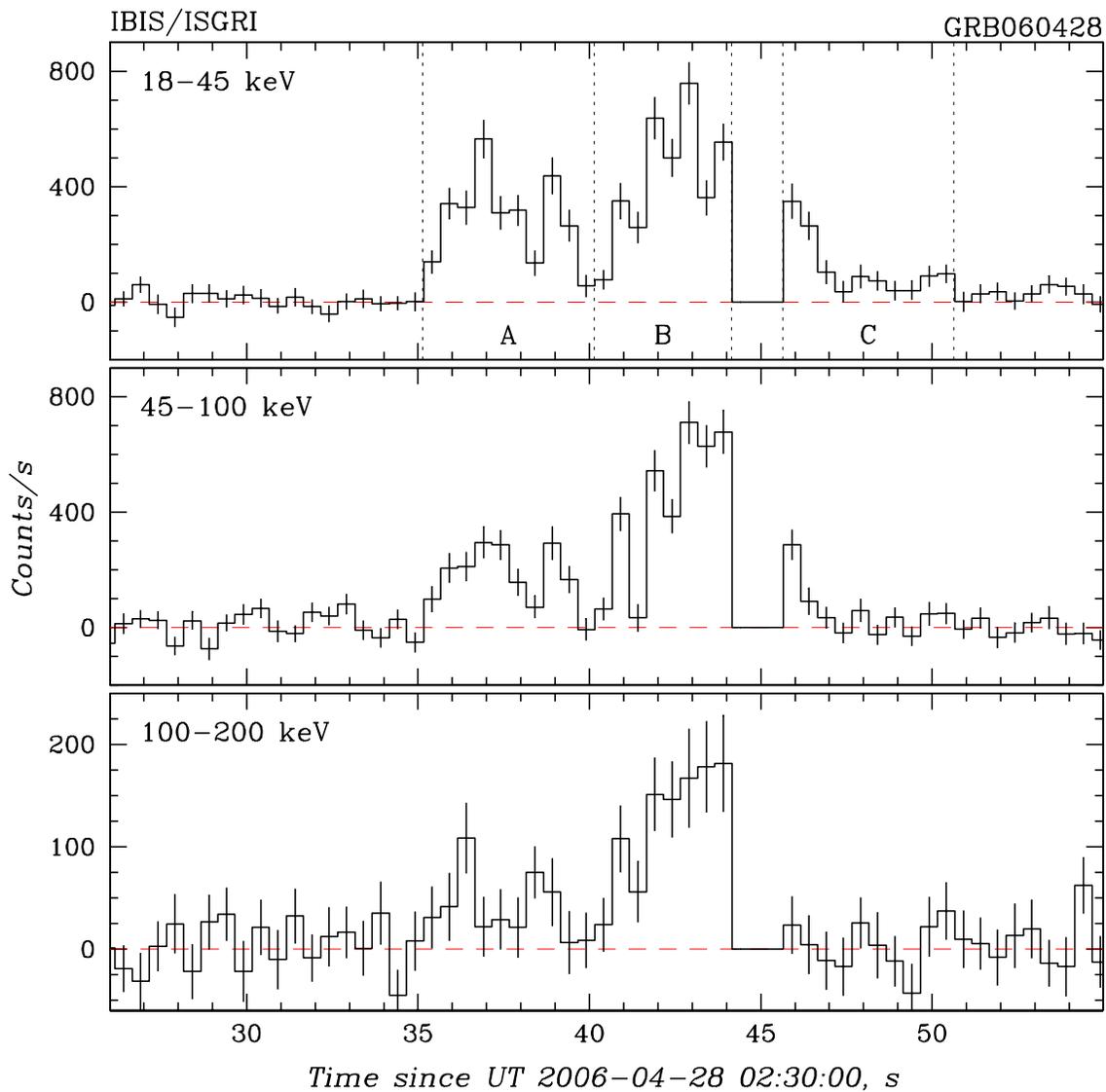,width=0.98\linewidth}

\caption{\rm Time profile of GRB\,060428C obtained by IBIS/ISGRI
in the energy bands 18--45, 45--100 and 100--200 keV. Time on
the X-axis is given in seconds after 2\uh30\um00\us\ UT, the
time bin is 0.5 s.\label{fig:lc_isgri}}
\end{figure}

It is obvious from the figure that the second flare was decaying
faster in the hard $\ga100$ keV energy bands of SPI, IBIS/ISGRI
and KONUS/WIND than in the softer ones. In this connection it is
interesting to trace the evolution of the burst's profile with
energy.  The IBIS/ISGRI profiles measured with a resolution of
0.5 s in the 18--45, 45--100 and 100--200 keV bands are shown in
Fig.\,\ref{fig:lc_isgri} from top to bottom. These profiles were
obtained with the same method as the profile in
Fig.\,\ref{fig:lc_isgri_0.2s}. The figure confirms the presence
of two flares (intervals A and B in the top panel of the figure)
in the profile of the main event as well as the presence of weak
remaining emission (interval C and the later observations). The
spectrum of the first flare was obviously softer than that of
the second flare. This means that the emission hardness has been
increasing during the main event, which is pretty unusual for
$\gamma$-ray bursts. It is likely that the figure indicates also
some smoothing of the short term variability in the burst profile
during the transition to higher energies. And finally, the
figure shows that the emission during the 
interval C, at the stage of early afterglow, again becomes
softer than that of the second flare of the main event, thus the
second flare indeed was gading faster at high $\ga100$
keV energies than at the low ones.
\begin{figure}[thp]
\centering
\epsfig{file=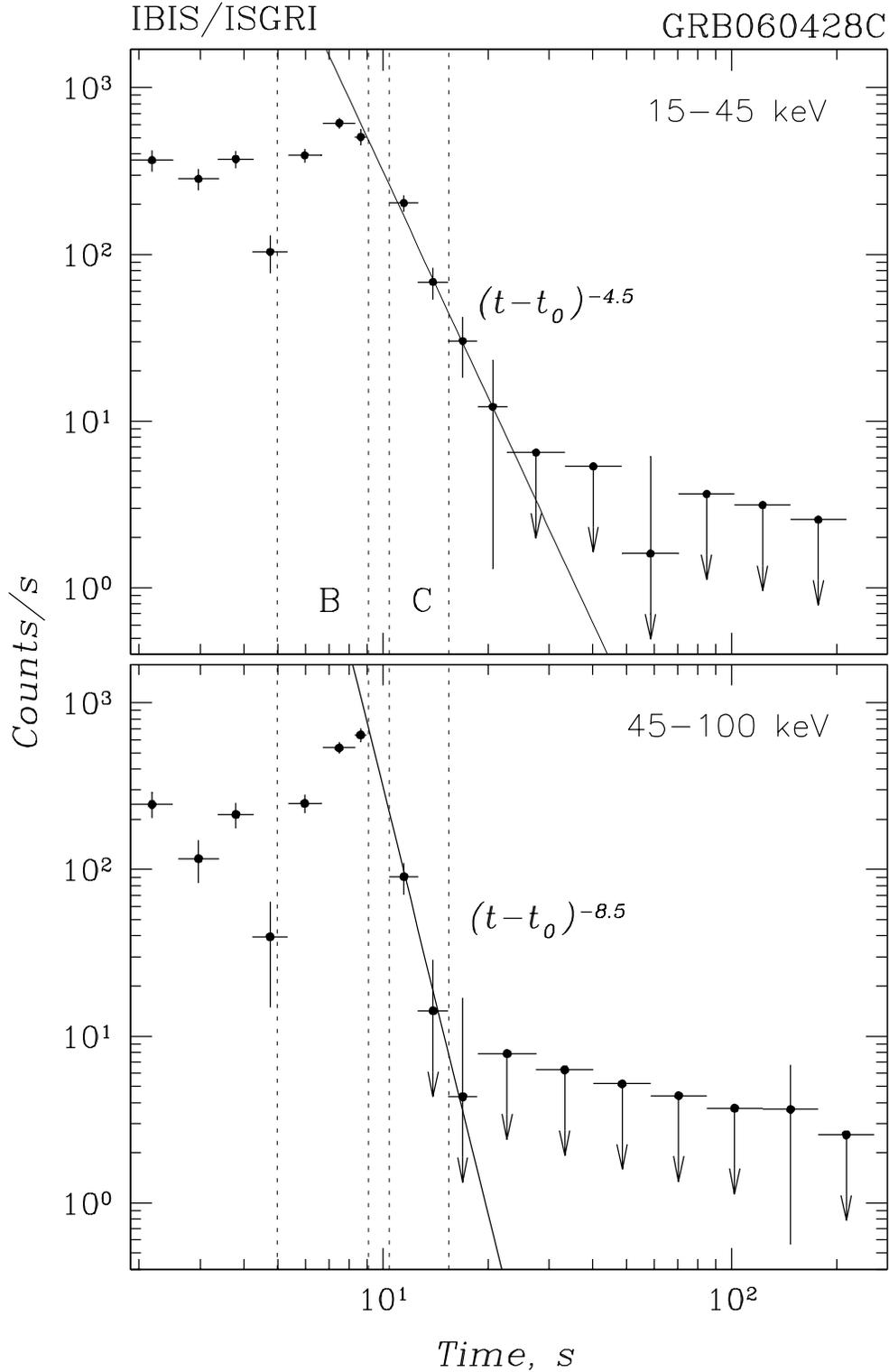,width=0.9\linewidth}

\caption{\rm Time profile of GRB\,060428C obtained by the
IBIS/ISGRI telescope in the energy bands 15--45 and 45--100 
keV. Time on the X-axis is given in seconds since the beginning
of the burst. The logarithmic axes are used to show a power law
like dependence of the afterglow on time. The rate of the
afterglow fading raises with energy.\label{fig:lc_isgri_log}}
\end{figure}
To demonstrate the fundamental difference of the emission at
stage C from that at stages A and B, we show in
Fig.\,\ref{fig:lc_isgri_log} the IBIS/ISGRI profiles of the
burst in different energy bands in a double logarithmic scale.
It is seen that the flux during stage C and right after it (up
to $\sim 30$ s) falls according to a power law with an index
$\gamma_1\simeq-4.5$ in the 15--45 keV energy band and
$\gamma_2\ga-8.5$ in the 45--100 keV band. The power law like
decay is characteristic of the $\gamma$-ray burst afterglow
observed at their early stage for example by Burenin et
al. (1999a,b), Tkachenko et al. (2000), Barthelmy et al. (2005),
Vaughan et al. (2006), Nousek et al. (2006), O'Brien et
al. (2006) and other authors.

\begin{figure}[t]

\centering
\epsfig{file=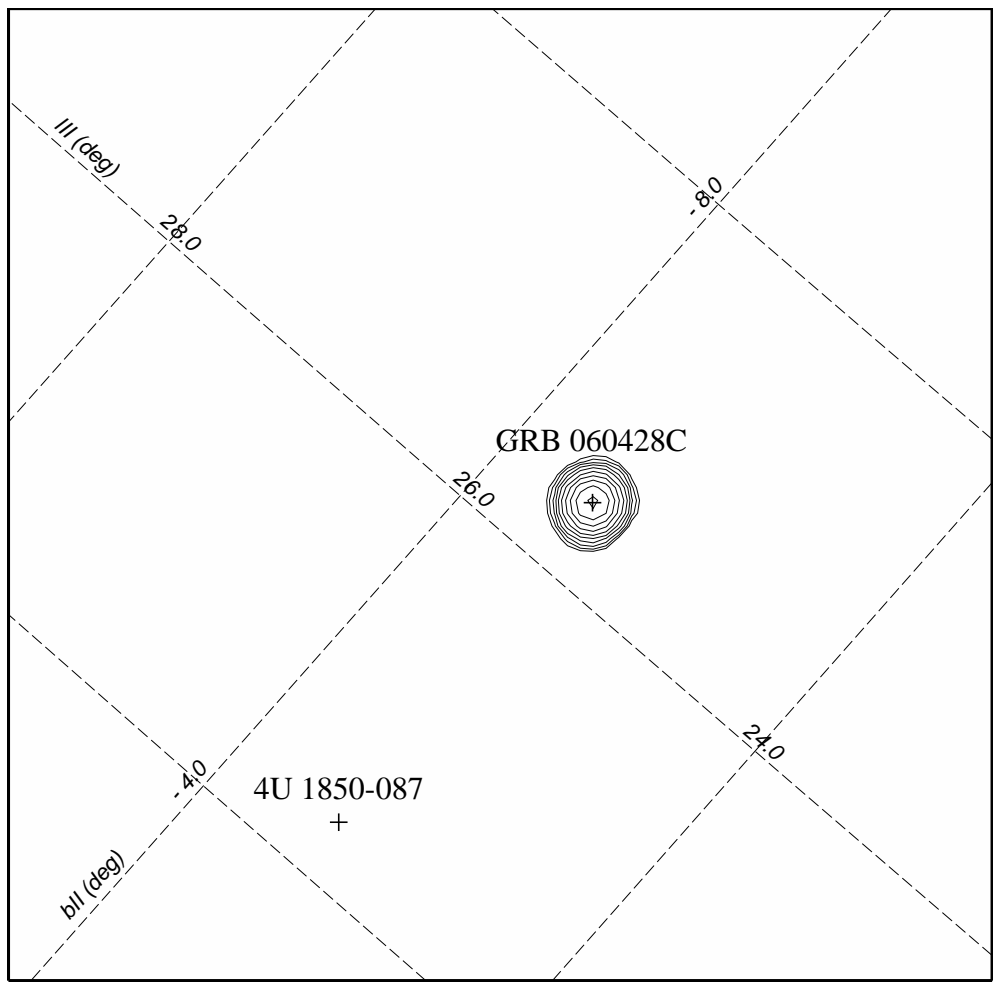,width=0.8\linewidth} 

\caption{\rm Image of the sky within the IBIS/ISGRI field of
view of $5\fdg0\times5\fdg0$ in size obtained during GRB\,060428C
(the exposure time is 13 s, the energy band is 20--200
keV). Contours denote regions of reliable detection of sources
at the signal to noise ratio $S/N=4.0, 5.2, 6.7, 8.6, 11.1,
..., 40$.
\label{fig:ima_isgri}}
\end{figure}

\section*{LOCALIZATION}

Figure \,\ref{fig:ima_isgri} shows an image (a map of the signal
to noise ratio $S/N$), obtained by the IBIS/ISGRI telescope in
the 20--200 keV band during the first 13 s of the
burst. Although there are many known X- and $\gamma$-ray sources
located in this region of the sky, none was detected at the
confident level during these 13 s. GRB\,060428C was detected in
the given energy band at the $S/N$ level $\simeq38$.  The
position of the burst's source, $R.A.=19\uh00\um52\us$,
$Decl.=-9\deg33\arcmin00\arcsec$ (equinox 2000.0) was determined
by the IBIS telescope with an accuracy of better than
2\arcmin. This accuracy notably exceeds ability of the
triangulation method even if data of the KONUS/WIND instrument
are used.


\section*{SPECTRUM OF THE BURST}
\begin{figure}[t]
\centering
\epsfig{file=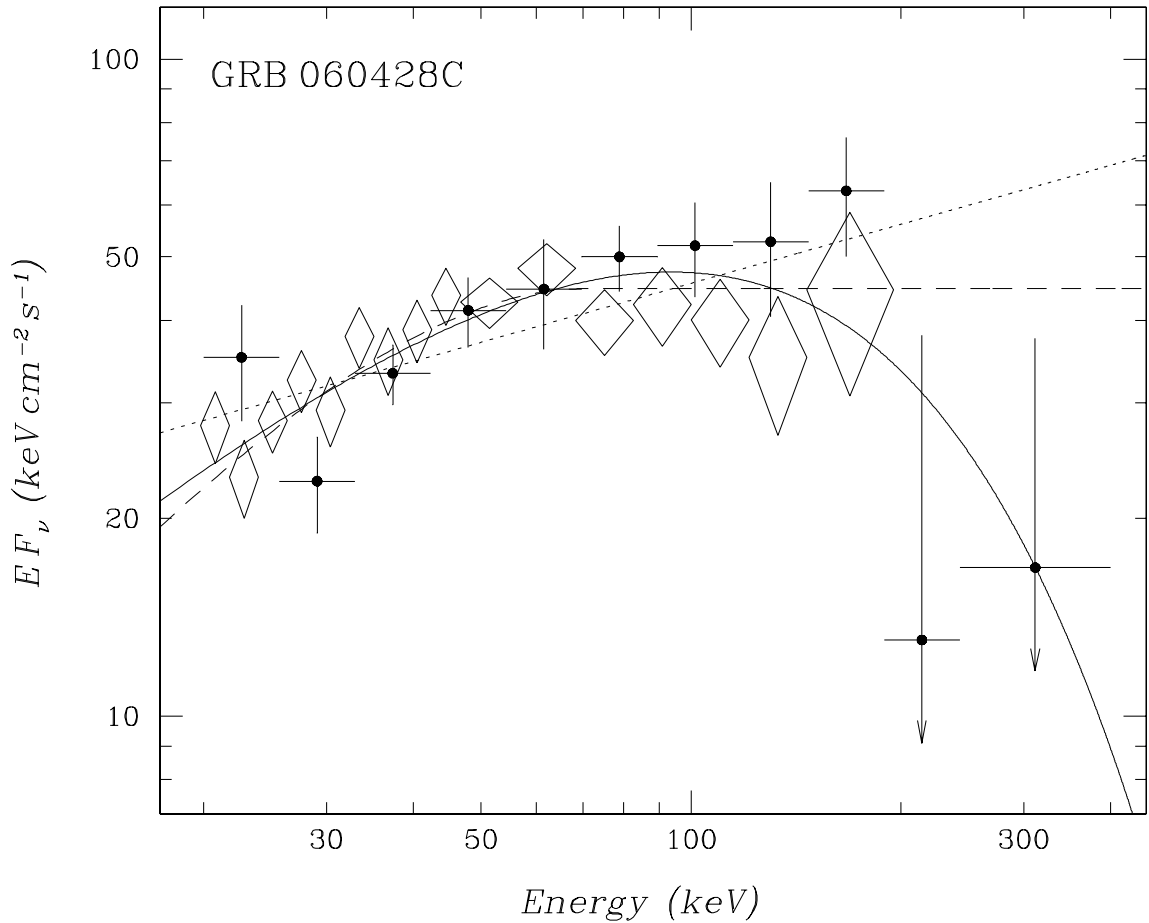,width=0.98\linewidth}
\caption{\rm Mean emission spectrum of GRB\,060428C measured
with INTEGRAL, by the instruments IBIS/ISGRI (diamonds) and SPI
(crosses), during the first $\sim15$ s of the burst and its
approximation by simple models: a power law (dotted line), a
power law with an exponential high energy cut-off (solid line),
and Band's model (Band et al. 1993) with the fixed spectral
slope at high energies $\beta=-2$ (dashed line).
\label{fig:sp_broad}}
\end{figure}

The mean IBIS/ISGRI and SPI spectrum of GRB\,060428C accumulated
during the first 15.5 s of the burst (during intervals A, B and
C indicated in the upper panel of Fig.\,\ref{fig:lc_isgri}) is
shown in Fig.~\ref{fig:sp_broad}. The SPI data allow ones to
trace the burst's spectrum up to $\sim 400$ keV, although the
IBIS/ISGRI data are more confident below $\sim 200$ keV. Table
~1 lists the results of approximation of this spectrum in the
20--400 keV band with simple analytical models: a power law (PL)
with a photon index $\alpha$, a power low with an exponential
cut-off (CP) defined by an energy $E_0$, and a double power law
model (BM) by Band et al. (1993), which is broadly used in
studies of $\gamma$-ray bursts. In Band's model, the high energy
power law index was frozen at the value $\beta=-2$. It is
obvious that the models with a high energy cut-off (CP and BM)
describe the observed spectrum much better than a single power
law. It is easy to find the position of a maximum in the energy
spectrum (spectrum in units of $E^2\,dN/dE$) for the CP model
--- this position determines the ``peak energy'' of the burst
$E_p=(2+\alpha)E_0\simeq 94$ keV.

\begin{table}[t]
\centering 
{{\bf Table 1.} Approximation of the GRB\,060428C
  spectrum from the IBIS/ISGRI and SPI data obtained in the
  20--400 keV band during the first $\sim15$ s of the
  burst.}\label{spbroad}

\small
\vspace{5mm}\begin{tabular}{c|c|c|c|c|c|c} \hline\hline
Model&$E_0$\a\ &$\alpha$\b\ &$\beta$\c\ &$F_{\rm X}$\d\ &$F_{\rm B}$\e\ &$\chi^2(N)$\f\  \\ \hline
PL   &$            $&$-1.70\pm0.05$&      &$1.50\pm0.04$ &$2.18\pm0.06$ &1.47 (34)\\
CP   &$102(-25/+43)$&$-1.08\pm0.18$&      &$1.45\pm0.04$ &$1.68\pm0.04$ &1.00 (33)\\
BM   &$ 54\pm14$&$-0.71\pm0.26$&$-2.0$ &$1.46\pm0.04$ &$1.96\pm0.05$ &0.98 (33)\\ \hline
\multicolumn{6}{l}{}\\ [-3mm]
\multicolumn{6}{l}{\a\ cut-off energy (keV)}\\
\multicolumn{6}{l}{\b\ photon index}\\
\multicolumn{6}{l}{\c\ photon index at high energies in the model by Band et al. (1993)}\\
\multicolumn{6}{l}{\d\ flux in the 20--200 keV band ($10^{-7} \
  \mbox{erg cm}^{-2} \ \mbox{s}^{-1}$)}\\ 
\multicolumn{6}{l}{\e\ flux in the 20--400 keV band ($10^{-7} \
  \mbox{erg cm}^{-2} \ \mbox{s}^{-1}$)}\\ 
\multicolumn{6}{l}{\f\ best-fit value of $\chi^2$ normalized for $N$
($N$ -- number of degrees of freedom)}\\
\end{tabular}
\end{table}
\begin{figure}[h]
\centering
\epsfig{file=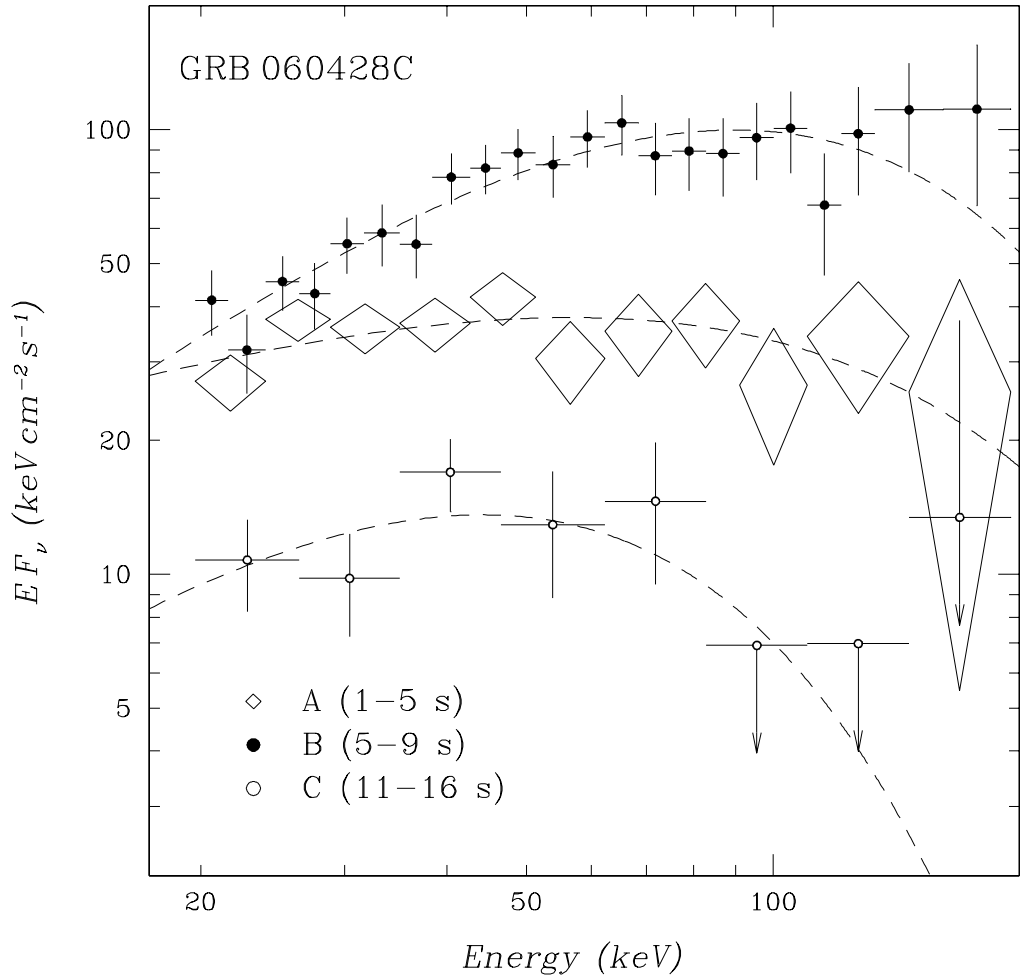,width=0.9\linewidth}
\caption{\rm Spectral evolution of the GRB\,060428C emission
according to IBIS/ISGRI (spectra A, B and C were accumulated
during the corresponding time intervals 
shown in the upper panel of Fig.\,\ref{fig:lc_isgri}).}
\label{fig:sp_evol}
\end{figure}

\normalsize

The evolution of the spectrum of GRB\,060428C is shown in
Fig.\,\ref{fig:sp_evol}.  The spectra A, B and C were obtained
by the IBIS/ISGRI telescope at different stages of the burst
(the time intervals corresponding to these stages are shown in
Fig.\,\ref{fig:lc_isgri}). Dashed lines show the best-fit
approximations of the spectra by a power law with an exponential
cut-off at high energies. The results of approximation of these
spectra by models PL and CP are listed in Table\,2. It is seen
that the emission spectrum was relatively soft (photon
index of the model PL $\alpha\sim-1.9$) at the initial stage of
the burst (during the first flare), it became to be much harder
($\alpha\sim -1.5$) at the end of the main event (during the
second flare), but softened again during the afterglow
($\alpha\sim -2.1$).


\section*{DISCUSSION}

Although GRB\,060428C has much of X- and $\gamma$-ray properties
in common with those of the typical $\gamma$-ray bursts (for
example, a hard almost power law spectrum extended to $\ga400$
keV, a peak energy $E_p\simeq100$ keV, peak
$F_p\simeq3.6\times10^{-7}$ erg cm$^{-2}$ s$^{-1}$ and
total $F\Delta T\simeq 2.3\times10^{-6}$ erg
cm$^{-2}$ fluxes, a duration of the main event $\Delta T\sim10$
s), it demonstrated simultaneously a number of features that made 
it to be very unusual and extremely interesting. In particular,
only a few bursts were known before its discovery which hardness
was increasing during the burst. There was of great interest a
complicated multi-peak profile indicating of powerful processes
of collisions of internal shock waves within the expanding
envelope of the burst's source. The duration of individual peaks
$\delta T\la 0.1$ s $\sim 10^{-2}\Delta T$ was defined by time
lasting between relativistic ejecta responsible for colliding
shock waves. But probably the most interesting feature of this
burst was its early afterglow, decaying according to a steep
power law with the index $\gamma\simeq-4.5$. Such a steep decay
may be explained within the model by Kumar \& Panaitesku (2000) of
high-latitude emission of the burst occurred in a low density
medium (the emission of relativistic shells moving with
large angles $\theta>\Gamma^{-1}$ to the directon to an observer, here $\Gamma$
is a Lorenz factor of the shell). In this model the law for
a decay of the soft X-ray emission $F_{\nu}\sim(T/\delta
T)^{\gamma}$ is determined by the equation
$\gamma=-2+(\alpha+1)\simeq-2.5$ where $\alpha\simeq-1.5$ is a
photon index of the hard emission of the main event. The observed
afterglow is a combination of the high-latitude emission of all
peaks in the profile. At higher energies, relativistic effects
lead to the shift of the cut-off in the spectrum towards the softer
band (the peak energy depends on time as $E_p\sim\delta
T/T$). The spectral softening observed with IBIS/ISGRI at the
afterglow stage of the burst relatively the main event, the
absence (weakening) of the afterglow in the hard energy channels
100--200 and 45--100 keV relatively the soft channel, and even
the faster decay of the afterglow in the soft band relatively
the theoretical dependency $T^{-2.5}$ may be a direct consequence
of these effects. Note that the peak energy decreases $\sim2$
times during the transition from the spectrum B to the spectrum
C (see Table\,2). The faster decay of the afterglow may take
place and in the case of strongly collimated gamma-ray burst.
\begin{table}[t]
\centering
{{\bf Table 2.} Approximation of the
 \mbox{GRB\,060428C} spectrum at different stages of its evolution.}\label{evolsp} 

\small
\vspace{5mm}\begin{tabular}{c|c|c|c|c|c} \hline\hline
Spectrum&Model&$E_0$\a\ &$\alpha$\b\ &$F_{\rm X}$\c\
&$\chi^2(N)$\d\  \\ \hline
A     &PL   &              &$-1.95\pm0.11$&$1.29\pm0.07$ &1.02 (22)\\
      &CP   &$ 92\pm40 $&$-1.39\pm0.12$&$1.20\pm0.06$ &0.97 (21)\\
&&&&&\\
B     &PL   &$            $&$-1.50\pm0.07$&$3.05\pm0.11$ &1.06 (22)\\
      &CP   &$ 61\pm19 $&$-0.52\pm0.08$&$2.77\pm0.10$ &0.56 (21)\\
&&&&&\\
C     &PL   &$            $&$-2.12\pm0.28$&$0.39\pm0.05$ &0.98 (22)\\
      &CP   &$ 29.8\pm6.4 $&$-0.52$\e\ &$0.32\pm0.04$&0.80 (22)\\ \hline
\multicolumn{6}{l}{}\\ [-3mm]
\multicolumn{6}{l}{\a\ cut-off energy (keV)}\\
\multicolumn{6}{l}{\b\ photon index}\\
\multicolumn{6}{l}{\c\ flux in the 20--200 keV band ($10^{-7} \
  \mbox{erg cm}^{-2} \ \mbox{s}^{-1}$)}\\ 
\multicolumn{6}{l}{\d\ best fit value of $\chi^2$ normalized for  
 $N$ ($N$ -- number of degrees of freedom)}\\
\multicolumn{6}{l}{\e\ frozen with the value obtained for the spectrum
  B }\\
\end{tabular}
\normalsize
\end{table}
 
The authors thank R.A. Burenin, S.Yu. Sazonov, K.A. Postnov and
A.S. Pozanenko for valuable discussions. This research was based
on data of the INTEGRAL observatory obtained through the Russian
and European Science Data Centers for INTEGRAL. The research was
supported by the Russian Foundation of Basic Research (through
grant 05-02-17454), by the Presidium of the Russian Academy of
Sciences (through the program ``Origin and evolution of stars
and galaxies'') and by the program of support of Russian
leading scientific schools (through grant NSh-1100.2006.2).

\pagebreak   

\end{document}